# The design of a time-interleaved analog-digital conversion modulator based on FPGA-TDC for PET application

Cong Ma, Wubin Wang, Xiaokun Zhao, Li Yu, Guocheng Wu

R&D department of molecular imaging detector

Minfound medical system Co. Ltd

Shaoxing city, China

macong@mail.ustc.edu.cn, stepine2019@163.com

*Abstract*— **Fully Field Programmable Gate Array (FPGA) based digitizer for high-resolution time and energy measurement is an attractive low cost solution for the readout electronics in positron emission computed tomography (PET) detector. In recent years, the FPGA based time-digital converter (FPGA-TDC) has been widely used for time measurement in the commercial PET scanners. Yet, for the energy measurement, few studies have been reported on a fully FPGA based, large dynamic range and high resolution alternative to the commercial analog-digital converter (ADC). Our previous research presents a 25 Ms/s FPGA-TDC based free-running ADC (FPGA-ADC), and successfully employed it in the readout electronics for PET detector. In this work-in-progress study, by means of the time-interleaved strategy, a 50 Ms/s FPGA-ADC is presented. With only two off-chip resistors, both the A/D conversion and energy measurement are achieved on a Xilinx Kintex-7 FPGA. Therefore, this method has great advantages in improving system integration. Initial performance tests are also presented, and we hope it can give us a possibility to develop a new FPGA-only front-end digitizer for PET in future.**

*Keywords—FPGA, PET, ADC, TDC, energy measurement*

## I. Introduction

A positron emission computed tomography (PET) detector module needs to measure the arrival time, energy, and position of the incident 511-keV photon [1]. To identify tiny scintillator elements for high spatial resolution, a PET scanner requires tens of thousands of readout channels. Power consumption and density become challenging issues when the readout electronics system scales up. Although Application Specific Integrated Circuit (ASIC) technology can be resorted to solve the problem, its long development cycle and high cost is not feasible for laboratory experiments [2]. Thus, in many applications, light-sharing technique based on a multiplexing network with discrete components is still a common choice to reduce the number of signal processing channels [3-4]. Even so, the number of readout channels is still very large for a general PET scanner. There is a strong demand for simple and efficient readout electronics to use in the development of a compact and cost-effective PET system.

Researches in recent years succeed in integrating high-resolution time-digital converters into the field programmable gate array device (FPGA-TDC), which makes the time measurement much cheaper and simpler [5]. Nevertheless, the energy measurement is commonly based on a commercial high speed analog–digital converter (ADC) chip to digitalize the analogue signals, which certainly leads to high system power consumption and cost. So far the most common alternative solution is time-over-threshold (TOT) method which compares the integral signal with a pre-set threshold to translate the signal amplitude into a TOT time width measured with TDC [6]. For some one-to-one readout or hybrid charge division systems, it can be used because the pulse shape does not change with respect to position. However, its drawbacks of nonlinearity, small dynamic range, and low SNR make it impractical for high precision measurement especially in light-shared systems. Another method is called multi-voltage threshold (MVT) for digitizing a PET event by sampling with respect to certain reference voltages. By fitting the mathematical model of SiPM signals to the several sampled pulses, a high energy resolution with low dead time can be obtained. Nevertheless, MVT circuit consumes lots of comparators or FPGA IOs [7]. Thus, it is still a great challenge to achieve high resolution energy measurement over a large dynamic range without an ADC chip.

We previously reported a 25 Ms/s FPGA-TDC based free-running ADC (FPGA-ADC), and successfully employed it in the readout electronics for both one-to-one coupled and light-shared PET detectors based on SiPM and LYSO arrays [8]. This simple FPGA-ADC based on a carry chain TDC implemented on a Kintex-7 FPGA consists of only one off-chip resistor and two FPGA IOs so it has greater advantages in improving system integration and reducing cost than commercial chips. However, due to the low sampling rate of these ADC, we have to employ an analogue shaper to widen the signals (~600 ns) from SiPMs, which certainly increases system complexity and deteriorates the system's dead time and pipe up performance . To optimized the sampling rate performance, by means of the time-interleaved strategy, this work-in-progress study presents a 50 Ms/s FPGA-ADC. With only two off-chip resistors, both the A/D conversion and energy measurement are achieved on a Xilinx Kintex-7 FPGA. Compared with the TOT and MVT circuit, this method does not need additional digital-analog converter (DAC) or other active device, and we can get the real-time waveforms of the analogue signals. The electronics evaluation test results are presented, and an initial experiment was also conducted using a single LYSO crystal under $^{22}$Na point source excitation. We hope it can give us a possibility to develop a new FPGA-only front-end digitizer for PET in future.

## II. PRINCIPLE AND CIRCUIT DESIGN

### A. Time-inteleaved FPGA-ADC modulator

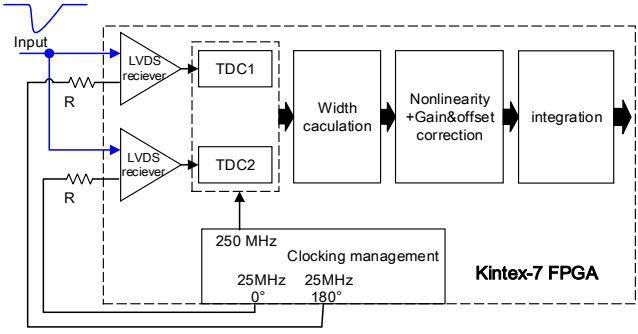

Figure 1 structure of the time-interleaved modulator

The structure of the presented time-interleaved FPGA-ADC is shown in Figure 1. Two clocks with a phase shift of 180 ° synthetized by the internal phase locked loop (PLL) are separately filtered to a quasi-triangular sampling ramp by a simple low pass filter formed by a serial resistors (Rs) with the parasitic capacitance (Cp) at the input of the user I/O ports. The analogue input signal is discriminated by the sampling ramps using the differential FPGA I/Os working in LVDS receiver mode and the width of the output digital pulses can be used to characterize the voltage amplitude of the analog signals. The discriminated widths are then measured by two following carry chain TDCs. Compared with the previous single sampling scheme, the time-interleaved modulator doubles the sampling rate. The high level of output clocks is 3.3 V.

On the basis of the precous research, the sampling frequency (Fs) of single sampling and Rs are configured as 25 MHz and 90 ohms. The circuit simulation indicates the measurable amplitude range can reach around 1.7 V (0.8 V–2.5 V). Besides, the employed TDC is normalized to 9 bits after online delay tap calibration (bin-by-bin) with an equivalent bin size of around 7.8 ps. Considering the maximum width of the measured pulses is 40 ns, the range of each single channel is larger than 12 bits.

### B. Correction methods

As mentioned before, the sampling ramps are not standard triangular ramps, which would lead to large conversion nonlinearity. This nonlinearity would bring harmonics at the multiples of the input frequency. To calibrate the conversation curve, we scanned different DC levels over the measurable amplitude range and made a look-up-table (LUT, size: 4096 × 4096) stored in FPGA to conduct a simple conversation nonlinearity correction mechanism.

Besides, for a time-interleaved A/D system, it is necessary to correct the gain, phase and offset difference between sampling channels. Generally, the correction algorithms are very complicated and resource intensive, which is not appropriate for the massive channels application [9]. Considering the sampling frequency of the FPGA-ADC is not very high, in this design we corrected only the gain and offset difference between the two time-interleaved channels. We reduced the phase difference by controlling the PCB layout and the TDC's routing constraints.

The ADC with online corrections has been implemented on a Kintex-7 FPGA (XC7K325T-2FFG900). The FPGA resource usage of one channel (including 2 TDCs with delay tap calibration, nonlinearity correction, gain&offset correction and the width calculation section) is listed in Table 1.

TABLE 1
FPGA RESOURCE USAGE OF ONE CHANNEL

|  | usage | Usage/total |
|---|---|---|
| LUT | 3960 | 1.94% |
| LUT RAM | 1726 | 2.7% |
| Block RAM | 5 | 1.12% |
| Flip-flop | 1418 | 0.35% |
| DSP | 1 | 0.12% |
| IO | 4 | 0.8% |
| Total on-chip power | 0.384 W | |

## III. INITIAL EXPERIMENT RESULTS

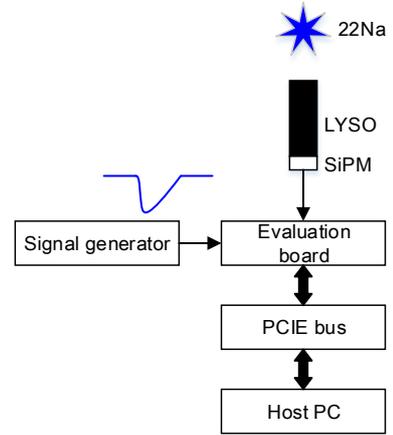

Figure 2 test platform

The test platform is shown in Figure 2. A signal generator is used to evaluate the ADC's electronics performance, and then a LYSO crystal (4 mm × 4 mm × 15 mm) coupled with a 4 mm × 4 mm SiPM (J-series from ON semiconductor) under $^{22}$Na excitation (~60 μi) is tested to get the single energy spectrum. The LYSO has been polished and later covered by reflective material (Teflon). The LYSO and the SiPM are pasted with optical grease (BC630). All the measured data, controlled commands, and debug information are communicated with the Host PC via PCIE bus.

## A. ADC's dynamic performance

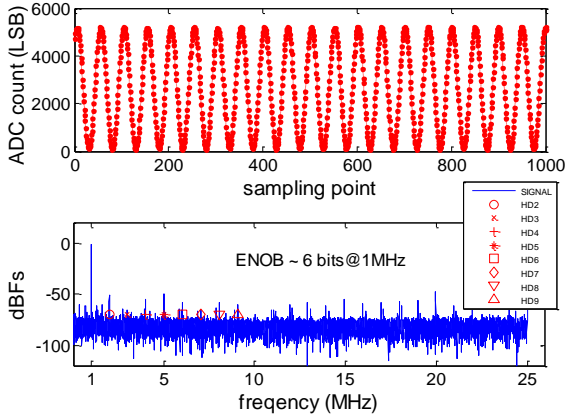

Figure 3 the ADC's dynamic performance (the frequency of input is 1 MHz)

The signal generator outputs continuous sine waves with a nearly full voltage swing to the ADC, and the dynamic performance is shown in Figure 3. The results indicate that the effective number of bits (ENOB) is around 6 bits at 1 MHz and 4.5 bits at 5 MHz.

## B. Measurment range and resolution

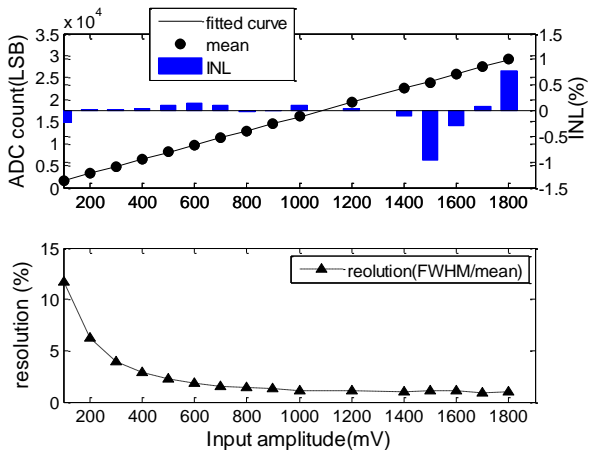

Figure 4 the measured conversation curve and resolution for the SiPM-like signals

In some light-shared PET applications, it is important to ensure good measurement linearity and resolution over a large dynamic range. We generated the signals according to the outputs of the SiPM's anode and the test results are shown in Figure 4. Test results indicate that the measurement integral nonlinearity (INL) is within ±1.0% without any correction and the resolution is better than 6.0% in the input range from 200 mV to 1.8 V with a 0.7 V DC offset.

## C. Energy spectrum

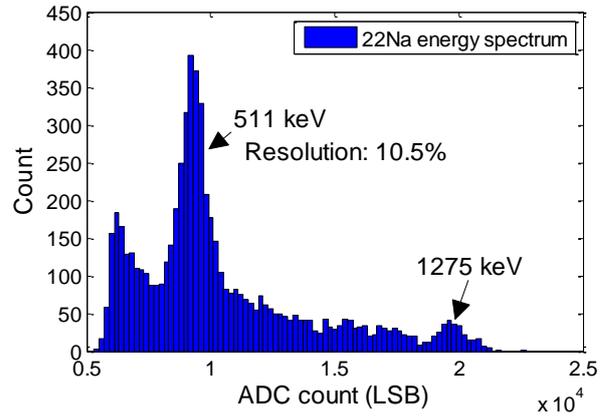

Figure 5 the tested $^{22}$Na energy spectrum

The tested energy spectrum of $^{22}$Na is shown in Figure 5. The test results indicate that the energy resolution is around 10.5 % at 511 keV before the SiPM saturation correction. The spectrum is almost the same as that acquired by the oscilloscope.

## IV. CONCULUSION AND FUTURE WORK

In this work-in-progress paper, we present a 50 Ms/s time-interleaved analog-digital conversion modulator based on FPGA-TDC. With only two off-chip resistors, all the A/D conversion and energy measurement are achieved on FPGA. Therefore, this method has greater advantages in improving system integration than traditional schemes. The initial test results indicate that this method can get a clear $^{22}$Na energy spectrum, which gives us a possibility to develop an FPGA-only front-end digitizer for PET in future.

In the next work, we will test the ADC's performance for the crystal array, especially in the light-shared application. And then a fully FPGA based digitizer for clinical PET scanner will be developed.